\documentclass[preprint,showpacs,showkeys,prb,aps]{revtex4}

\usepackage{graphicx}
\usepackage{dcolumn}
\usepackage{bm}
\usepackage{amssymb}
\usepackage{amsmath}

\begin{document}

\thispagestyle{empty}

\title{Knots in Macromolecules in Constraint Space}

\author{Michael Brill}
\author{Philipp M. Diesinger}
\author{Dieter W. Heermann}
\email{heermann@tphys.uni-heidelberg.de}
\homepage{http://wwwcp.tphys.uni-heidelberg.de}
\affiliation{
           Institut f\"ur Theoretische Physik \\
           Universit\"at Heidelberg \\
           Philosophenweg 19 \\
           D-69120 Heidelberg\\
           and\\
           Interdisziplin\"ares Zentrum\\
           f\"ur Wissenschaftliches Rechnen\\
           der Universit\"at Heidelberg}
\date{\today}%

$\\\\$

\begin{abstract}
We find a power law for the number of knot-monomers with an exponent $0.39 \pm0.13$ in agreement with previous simulations. For the average size
of a knot we also obtain a power law $N_m=2.56\cdot N^{0.20\pm0.04}$. We further present data on the average number of knots given a certain
chain length and confirm a power law behaviour for the number of knot-monomers. Furthermore we study the average crossing number for random and
self-avoiding walks as well as for a model polymer with and without geometric constraints. The data confirms the $aN\log N + bN$ law in the case
of without excluded volume and determines the constants $a$ and $b$ for various cases. For chains with excluded volume the data for chains up to
$N=1500$ is consistent with $aN\log N + bN$ rather than the proposed $N^{4/3}$ law. Nevertheless our fits show that the $N^{4/3}$ law is a
suitable approximation.
\end{abstract}


\maketitle

\section{Introduction}

Knots are of biological interest because they preserve topological information. In DNA packing and unpacking, an enzymatic reaction converts DNA
strands to knots. Such knots are actively removed under energy consumption from ATP, by topoisomerase II by cutting a DNA segment, passing
another segment through the gap, and resealing the cut of the former, until eventually the topology of the chain is that of an unknot. In
bacteria DNA occurs knotted, i.e. in a state topologically different from a simply connected ring. This can be seen using electron microscopes.
Underlying and overlying segments are distinguished using a protein coating. The flattened DNA is then visualized as a knot. The unknotting
number and ideal crossing number can then be estimated. For several DNA fragments with the same knot number there may be a variety of different
forms as the DNA is twisted and distorted. However, the average writhe and crossing number can be estimated for a particular ideal knot number.
A more convenient procedure for determining the crossing number of DNA knots involves using gel electrophoresis.

It has been shown that DNA knots having the same molecular size increase their speed of electrophoretic migration with increasing number of
nodes, i.e. the intersections of DNA segments in planar projections~\cite{Sundin}. However, these early gel systems did not succeed in
separating different DNA knots with the same minimal crossing number. It is a key question to understand the statistical behaviour of such
knotted DNA to understand a number of physiological processes having to overcome this knottedness, or to quantify results from DNA separation
techniques such as electrophoresis, in which the knottedness influences the mobility.

DNA in restricted volumes shows knots, for example when linear double-stranded DNA is packed inside bacteriophage capsids. For such a situation
the knotting probabilities of equilateral polygons confined into spherical volumes were calculated by Monte Carlo simulations~\cite{Arsuaga}.

Already a random walk (the DNA might be considered in some circumstances as a random walk)
can frequently lead to the formation of knots and it was conjectured
(Frisch-Wasserman-Delbr\"uck conjecture~\cite{del,Frisch}) and proven that as the walk
becomes very long the probability of forming nontrivial knots upon closure of such a walk tends
to one~\cite{Summers,Diao1,Pippenger}. In thermally fluctuating long linear polymer chains in solution, the ends
come from time to time into a direct contact or at close vicinity of each other. At such an instance,
the chain can be regarded as a closed one and thus will form a knot. Simple knots show their highest
occurrence for shorter random walks than more complex knots~\cite{Deguchi,Dobay}.

How does one identify a knot in a polymer chain? From a mathematical point of view one uses the Alexander-Jones polynomial or other types of
invariants, known as knot (or link) polynomials and defined via skein relations~\cite{Vologodskii1,Vologodskii2,Koniaris,Deguchi1,Deguchi2},
i.e. a set of rules defining a knot polynomial invariant. The Alexander polynomial is calculated from a two-dimensional projection of a
three-dimensional knot. It provides an invariant in as far as all the projections yield the same polynomial. Unfortunately, there are pairs of
distinct knots that share the same Alexander polynomial. This is typical in as far as the invariants in general are not one-to-one. Since there
is no exact algorithm for classifying all knots, in this paper we use a practical approach to the identification of knots. We apply a force to
the polymer stretching it. If there is a knot in the chain, it will tighten and can then be identified.

In this paper we take up the question of the statistics and properties of polymers with topological
and geometric constraints, in particular those with knots. From a theoretical point of view, the statistical
mechanics of such entangled systems is an unsolved problem\cite{Edwards1,Edwards2}.

As pointed out above, an important concept for the identification of knots and the statistics of them is the average crossing number (ACN). Diao
suggested to use the ACN as a measure of entanglement to determine whether a polymer chain (closed) is highly or weakly knotted~\cite{Diao1}.
For a given linear closed polymer the crossing number associated with a particular projection of the random walk is the number of crossings one
observes when the polymer is projected to a plane under the given projection direction. The average crossing number of the polymer is then
defined as the average of this crossing number over all possible projection directions~\cite{Diao1}.

We note that although a linear DNA behaves like a Gaussian walk up to rather high
lengths~\cite{Marko}, at fixed topology even a Gaussian chain behaves like an SAW-chain~\cite{Deutsch}.
Thus we investigate the knots or rather in the corresponding section the average crossing number
(ACN) for the case of phantom chains and chains with excluded volume. Both are investigated
with and without the constraint of presence of a wall excluding one half space.

For equilateral and for Gaussian random walks Diao~\cite{Diao1} succeeded in showing that the
average crossing number behaves $N\log N +cN$. For self-avoiding chains no analytical result is yet available
but a suggestion by Buck~\cite{Buck} that with excluded volume the behaviour changes to a power law.

In this paper we first consider the average crossing number for the case of random and self-avoiding walks under the constraint that they are
attached to a wall. Then we focus on the knot statistics and present results for the statistics of knots. We investigate the average number of
the knots given a certain chain length and confirm a power law behaviour for the number of knot-monomers.

\section{The average crossing number}

We start by investigating the effects of excluded volume interactions on the average crossing number (ACN) of equilateral and Gaussian random
walks (phantom chains) which represent polymer classes. To understand the effect of the excluded volume interaction, let us focus our attention
initially on the invariant

\begin{equation}
a(l_1,l_2) = \frac{1}{2 \pi} \int_{\gamma_1} \int_{\gamma_2}
 \frac{|\dot{\gamma}_1(t),\dot{\gamma}_2(s),\gamma_1(t)-\gamma_2(s)|}{\|\gamma_1(t)-\gamma_2(s)\|^3} \; dt \; ds
 \label{eq:acn1}
\end{equation}

\noindent which is the basis for the prediction~\cite{Diao_gauss,Diao_equi}
\begin{eqnarray}
\mbox{ACN}_{\mbox{Gaussian}}(N) &= \frac{1}{2\pi} N \; log(N) + O(N) \approx \frac{1}{2\pi} N \; log(N) + c_1 \cdot N\\
\mbox{ACN}_{\mbox{Equilateral}}(N) &= \frac{3}{16} N \; log(N) + O(N)\approx \frac{3}{16} N \; log(N) + c_2 \cdot N
 \label{eq:acn2}
\end{eqnarray}

\noindent for the phantom and equilateral chains of length $N$ without excluded volume. In the above $l_1$ and $l_2$ are segments of the chain
and $\gamma$ is the arclength parametrization. Eq. (\ref{eq:acn1}) is a link invariant which specifies the topological state of the polymer
which, however, is not of the one-to-one type~\cite{Kauffman,Adams} and is not a true invariant for knots~\cite{Vologodskii1,Calugareanu}. While
the above invariant does not succeed in uniquely characterizing the knot, it is the first element in a hierarchy.

In~\cite{Diao_gauss,Diao_equi} it was shown that for two chain segments $l_1, l_2$ on average $a(l_1,l_2)$ behaves as

\begin{equation}
<a(l_1,l_2)> \quad = \frac{1}{2 \pi d^2}+O\left( \frac{1}{d^{2.5}}\right )
\label{eq:e1}
\end{equation}

for Gaussian phantom and

\begin{equation}
<a(l_1,l_2)> \quad = \frac{1}{16d^2}+O\left(\frac{1}{d^3}\right)
\label{eq:e2}
\end{equation}

\noindent for equilateral phantom chains, where $d$ is the distance between the two considered chain segments. So far
no prediction has been derived for the case of chains with excluded volume interaction. Hence it is important to ask
how far the estimate also applies to the case of excluded volume interaction and how differences manifest themselves.

Using Monte Carlo simulation~\cite{Heermann,Binder-Heermann} we calculated the average crossing number
by counting the crossings in numerous projections of $\gamma$
and taking the average over all these crossing numbers. For every calculated average crossing number, we
averaged over $1000$ randomly chosen planes to obtain a good
estimate of the actual ACN value.

Four types of chains have been investigated by us: Gaussian and equilateral chains with and without excluded volume. All chains are open and
start at the origin. The excluded volume chains were generated using a Pivot-Algorithm, which for example can be found in \cite{MadrasSokal}. We
used a hard-core excluded volume potential, in order to speed up the simulations. Between two consecutive chain points there is an ellipsoidal
hard-core excluded volume. When generating chains with excluded volume interactions the Pivot-Algorithm simply rejects all chain conformations
which have at least two overlapping ellipsoids.

\begin{center}
\begin{figure}[ht]
\includegraphics[width=10cm]{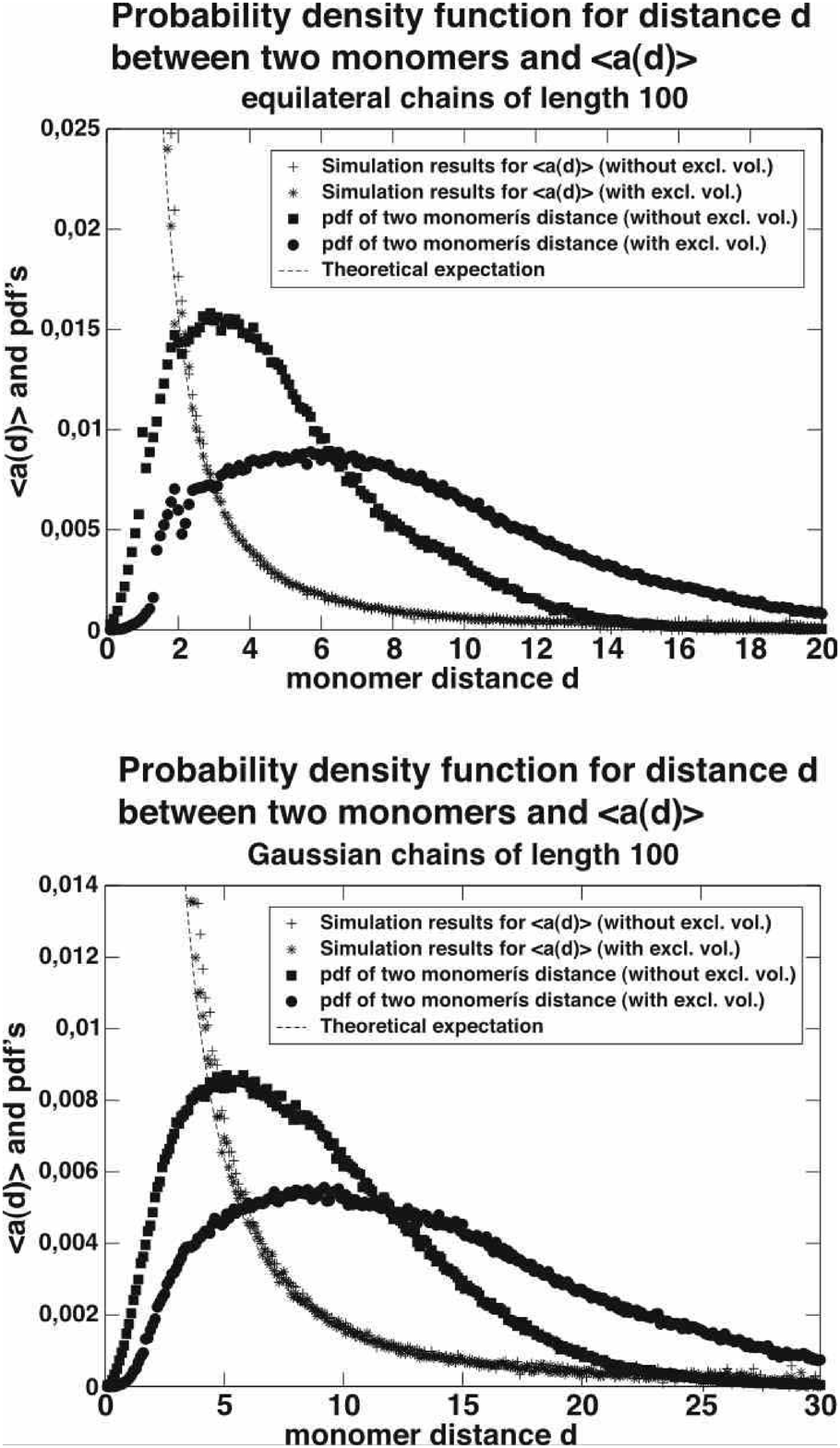}
\caption{\label{fig:acn2} Shown are the distance probability density functions (pdf) for Gaussian and equilateral chains with the cases of non-
and excluded volume. In both cases the distribution is rather different but yield the same behaviour on average (c.f. Figure~\ref{fig:acn1}). }
\end{figure}
\end{center}

\begin{center}
\begin{figure}
\includegraphics[width=\textwidth ,angle=0]{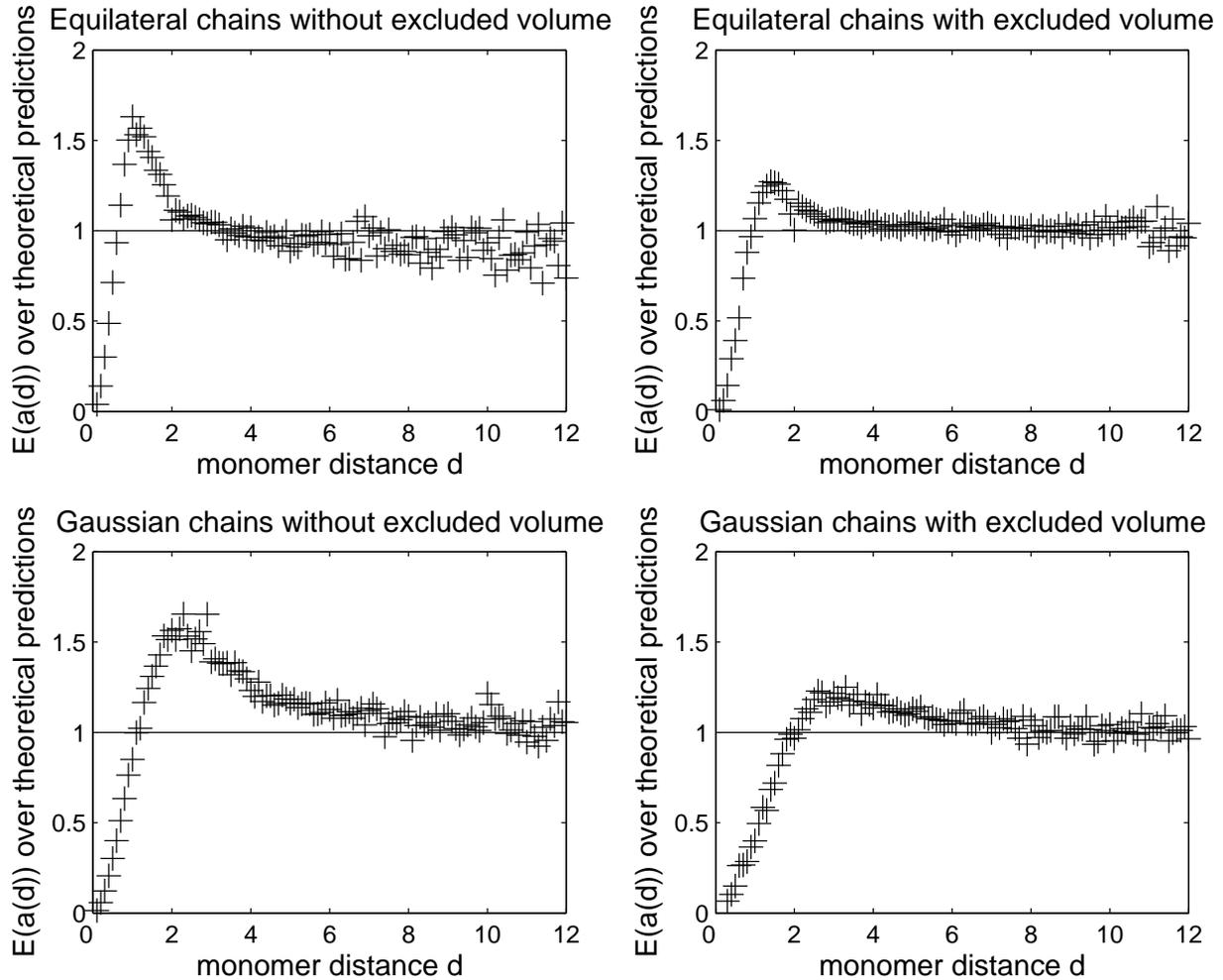}
\caption{\label{fig:acn1} Shown are the ratios between the prediction for the leading order term for the invariant $a(d)$ characterizing a knot
and the simulation results. While the prediction pertains to chains without excluded volume the results show the agreement with the theoretical
predictions are excellent for distances larger than $10$ for chains with and without excluded volume. The strong fluctuations at the end for the
larger distances are a consequence of the plotted ratio. A single point in the figure represents at least averages over 10 000 simulation
results.}
\end{figure}
\end{center}

\begin{center}
\begin{figure}[ht]
\includegraphics[width=\textwidth ]{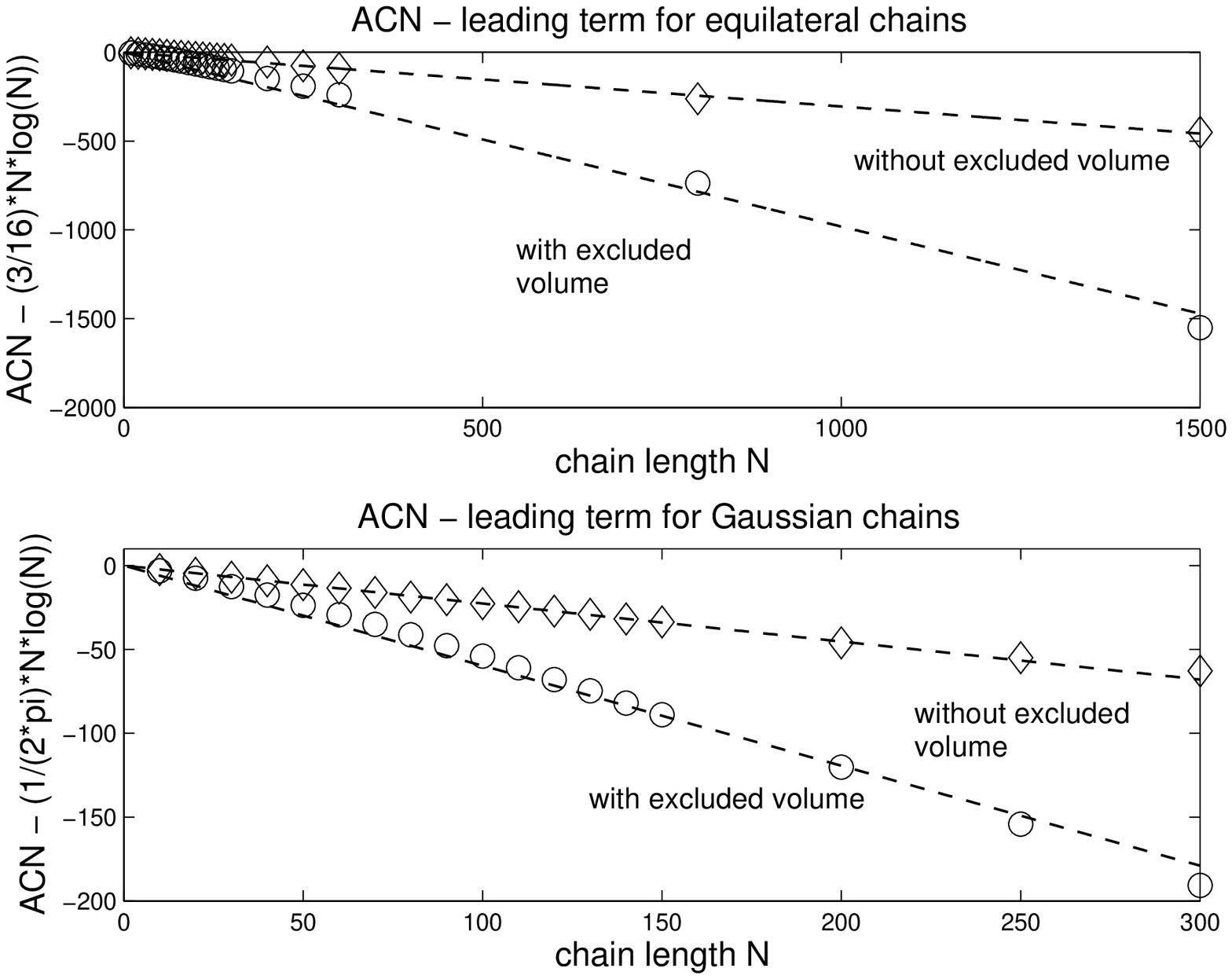}  
\caption{\label{fig:ACN2-lo} The error term of the theoretical prediction of the mean average crossing number is negative and much smaller for
the chains with excluded volume.}
\end{figure}
\end{center}

\begin{center}
\begin{figure}[ht]
\includegraphics[width=\textwidth ]{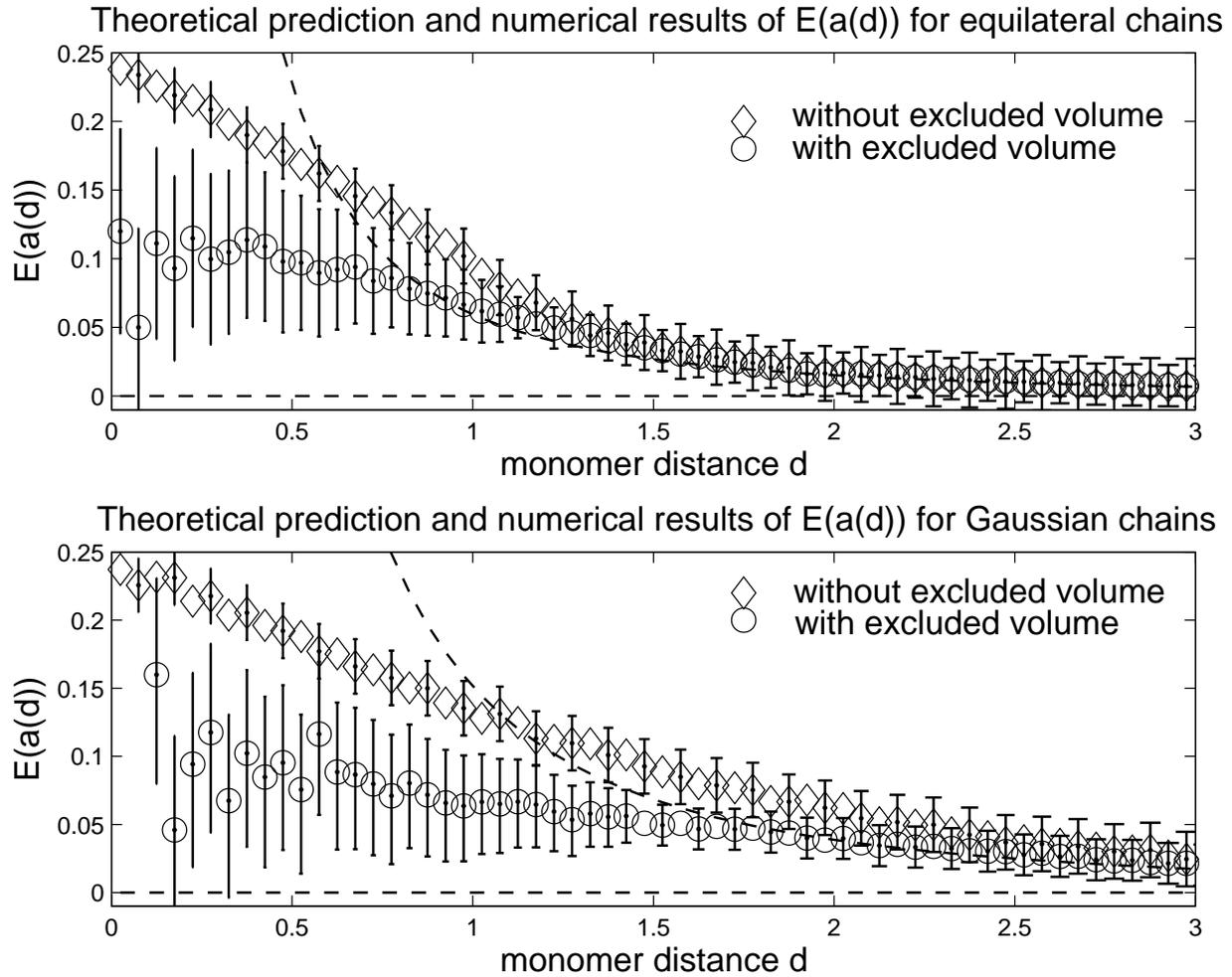} 
\caption{\label{fig:ACN3} The theoretical prediction (black line) and the numerical results for $\mathbb{E}(a(d))$ of Gaussian and equilateral
chains for small values of $d$. One can roughly see that $\mathbb{E}(a(d))$ is discontinuous at $d=0$ since $\mathbb{E}(a(0))=0$ (cf.
\cite{Diao_equi,Diao_gauss}).}
\end{figure}
\end{center}

We find that the excluded volume interactions do not have any measurable effect on the behaviour of $<a(l_1,l_2)>$ for distances larger than
$10$ chain segments, and only a slight one for shorter distances (see Figures~\ref{fig:acn2} to \ref{fig:ACN3}). This shows that there are
nearly no orientation effects on $<a(l_1,l_2)>$ due to the excluded volume interactions and mainly a dependence on the distance distribution of
the chain segments.

The strong reduction of $E(a(l_1,l_2))$ for chains with excluded volume (see Figure~\ref{fig:ACN2-lo}) is a result of the altered distance
probability density function (pdf) and not of any orientation effects. One can see that the chains with excluded volume are much more stretched
as expected due to the enhanced value of $\nu$ for the radius of gyration than those without. As a consequence the total $<a(l_1,l_2)>$ for the
chains with excluded volume is much lower than for chains without excluded volume.

The pdf of $d$ is shown in Figure ~\ref{fig:acn2} for equilateral and Gaussian chains of length $N=100$ with and without excluded volume. One
can see that the chains with excluded volume are much more stretched than those without and that the Gaussian chains are longer than the
equilateral ones. The latter is a consequence of the Gaussian probability distribution since the mean length of Gaussian chain segments is about
$1.6$ and the length of all equilateral chain segments is normalized.$\\$The pdfs of the equilateral chains show peaks at $d\approx 2$. This is
a feature of the equilateral chains since the distance of the endpoints of two consecutive line segments is always larger than 0 and smaller
than 2. If $d$ exceeds 2 the probability to find monomers with a distance $d$ drops immediately. In the case of the equilateral chains with
excluded volume the end to end distance of two consecutive line segments is always larger than
\begin{equation*}
d_{\text{min}}=\sin\left(\frac{\alpha_{\text{min}}}{2}\right)=0.66 \quad \text{with }\alpha_{\text{min}}=83.62°.
\end{equation*}
This can be seen in Figure~\ref{fig:acn2}, too: The pdf of the equilateral chains with excluded volume has two discontinuity points (at $
d_{\text{min}}=0.66$ and $d=2$). The one without excluded volume shows only the one at $d=2$.

 Of course the short range excluded volume effects are the larger ones. In this case $\mathbb{E}(a(d))$ is lowered because more
orthogonal orientations are more likely (cf.\ Figure~\ref{fig:ACN3}). But there are small long range excluded volume effects, too: Since the
orientation of two line segments within a chain with excluded volume slightly depends on all other segments of the chain the excluded volume
chains have another pdf as those without excluded volume and this means that the relative orientation of two line segments is altered too. But
this effect is very small and can only be seen for equilateral chains (cf.\ Figure~\ref{fig:acn2}). Since the Gaussian chain
segments have a distributed length, orientation effects due to long range excluded volume interactions could not be found in this case.

The fact that the ACN via $<a(l_1,l_2)>$ mainly depends on the distance probability density function $p_N(d)$ and to a much lesser degree on
orientation effects can be used to give a rough approximation for the ACN. The number of crossings of a chain of length $N$ is given by

\begin{eqnarray}
C(N) &=& \sum^{N-2}_{i=1} i - (N-2) = \frac{(N-1)}{2} \;(N-2) - (N-2) \\
     &=& \frac{N^2}{2}-\frac{5}{2}N+3
\end{eqnarray}

Now one can use the distance probability density function $p_N(d)$ as a weight for $<a(d)>$ to obtain an approximation for the ACN

\begin{equation}
\mbox{ACN} \approx \left ( \int_o^{\infty} <a(x)> \; p_N(x) \; dx \right ) \cdot C(N)
\end{equation}

As these random walks are a model for equilibrated polymers in solution and as these polymers do have excluded volume interactions one should
expect a much lower ACN for these polymers than predicted by \cite{Diao_gauss,Diao_equi}.

The prediction for the ACN by Diao et. al.~\cite{Diao_gauss,Diao_equi} as stated in Eq.(~\ref{eq:acn2}) are in good agreement with our
simulations. Our simulation results confirm these results and we calculated a factor of $c_2= -0.3051$ for the
equilateral chains and a factor of $c_1= -0.2265$ for the Gaussian ones.

Furthermore we found a $N \; log(N) + cN$-behavior for the chains with excluded volume, too rather than the proposed~\cite{Buck} $N^{4/3}$
law. The fit results are complied in the following two tables.
\begin{center}\begin{tabular}{|c|c|c|c|c|}
\hline \multicolumn{5}{|c|}{chains without excluded volume}\\ \hline  data&fit& $a$ & sse & rsquare\\ \hline
equilateral&$(3/16)n\log(n)+an$&-0.3051&445.2619&0.9998\\ \hline Gaussian&$(1/2\pi)n\log(n)+an$&-0.2265&61.38&1.0000 \\ \hline
\end{tabular}
\end{center}

\begin{center}\begin{tabular}{|c|c|c|c|c|}
\hline  \multicolumn{5}{|c|}{Gaussian chains with excluded volume}\\ \hline
 fit& $a$ &$b$& sse & rsquare\\
\hline $an^b$&0.03239&1.376&11.6763&0.9986\\ \hline $(1/2\pi)n\log(n)+an$&-0.5968&-----&537.53&0.9349\\ \hline
$an\log(n)+bn$&0.07468&-0.1553&16.3505&0.9980 \\ \hline $ax^{4/3}$&0.0407&-----&10.07&0.9988\\ \hline
\multicolumn{5}{|c|}{equilateral chains with excluded volume}\\ \hline $an^b$&0.06382&1.232&1306.1&0.9952\\ \hline
$(3/16)n\log(n)+an$&-0.9812&-----&29987&0.8896\\ \hline $an\log(n)+bn$&0.03914&0.05466&241.58&0.9991
\\ \hline $ax^{4/3}$&0.03086&-----&3606.5&0.9867\\ \hline
\end{tabular}
\end{center}

\section{Knots in macromolecules}

\begin{center}
\begin{figure}[ht]
\includegraphics[width=\textwidth]{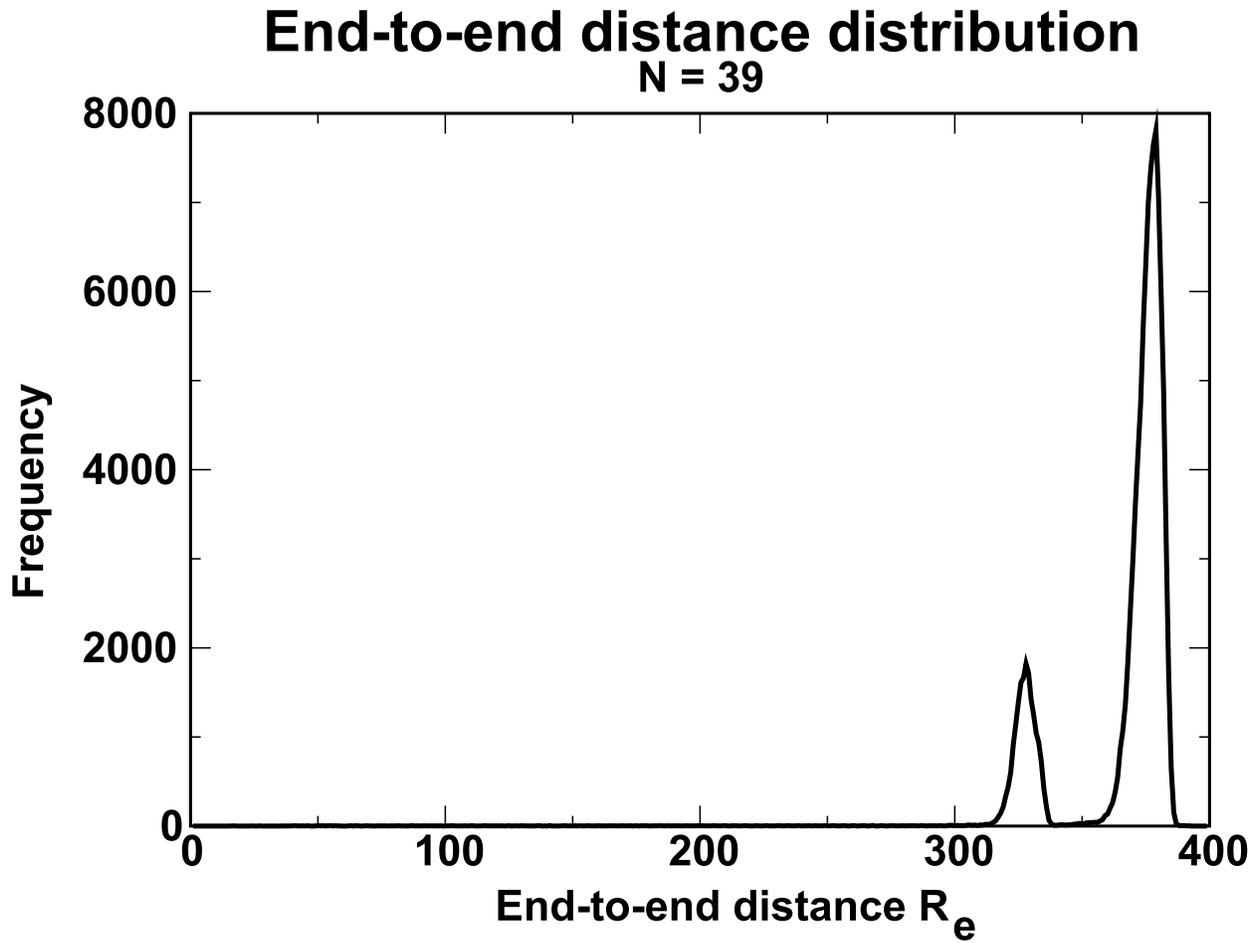}
\caption{\label{fig:e2edist} Distribution of the end-to-end distance of chains with and without knots. The plot shows the result of a sample of
ten independently generated chains and further sampling was done on these chains for the end-to-end distances. Chains with knots have end-to-end
distances in the left peak.}
\end{figure}
\end{center}

In this part, we focus on the knots in the chains. When one pulls on one end of a polymer chain with the other end fixed, at a wall for example,
it will be stretched. If the force is high enough, the end-to-end-distance distribution should have a large peak at over $95\%$ of the backbone
length. This is the case if the polymer is unknotted, i.e. when it is possible to stretch the chain completely. But if there are one or more
knots in the polymer chain, a part of the total possible length is lost! In this case the peak in the end-to-end distribution will be displaced
to smaller elongations. Shown in Figure~(\ref{fig:e2edist}) is the end-to-end distribution averaged over ten independently generated chains of
length of $N=39$. These chains were sampled for their end-to-end distribution during Molecular Dynamcis~\cite{Heermann} runs.

One can see two peaks, the larger one is the peak from the unknotted chains in the sample, the smaller one is the peak from the knotted chains.
The knots that occurred in these chains were later analysed to be simple ``trefoil'' knots.

In principle knots are curves with specific properties like average crossing number (as investigated in the previous section) and topology. Here
we apply a heuristic approach to the identification of a knot. We employ the notion of polymers with entanglement which can not be reduced to a
straight polymer chain. This means that the knots are an effect of the self-avoiding property of the macromolecules and they do only occur when
we stretch the chain. We do not want to investigate the topological properties of these chains but rather want to investigate their statistical
properties.

First let us see at many knots in polymer chains of a certain length can be found and how many monomers form such knotted places in the chains.
The chains we generate can only be in one half space restricted by an infinite wall. At this wall our polymers are fixed with one end, the other
end is used to pull on it. To create a starting configuration, we begin with a self-avoiding random walk~\cite{Binder-Heermann} with the origin
at the wall. After this we pull on the free end of the chain. What we obtain is a configuration that looks like a pearl necklace, with the
pearls being the entangled regions. We did not use Alexander-polynomials for the knot detection since our focus is on properties like the knot
size.

\begin{center}
\begin{figure}[ht]
\includegraphics[width=\textwidth]{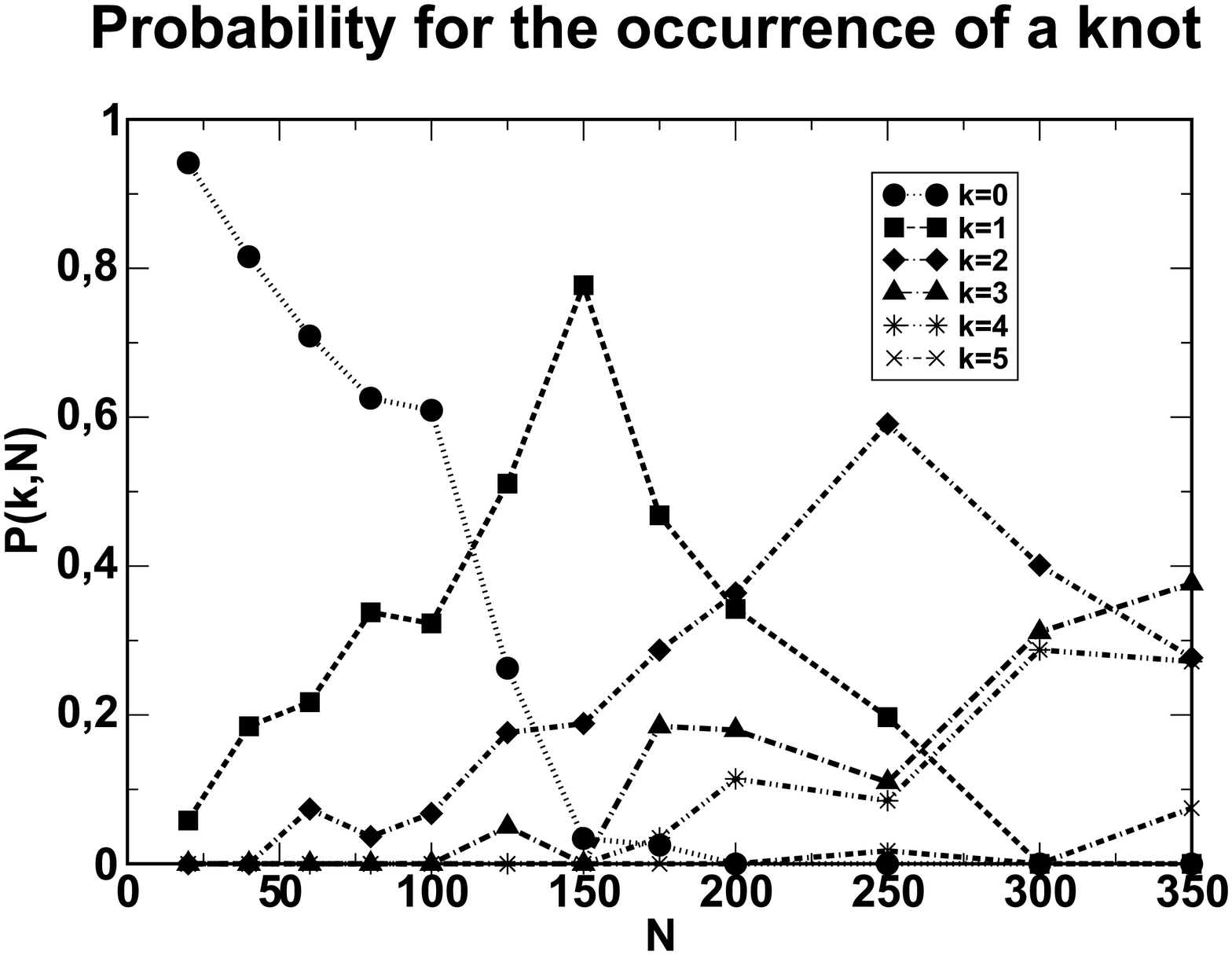}
\caption{\label{fig:knotprob350} Shown is the complement of the knot-probability for chains with up to 350 monomers. Shown is that data for
chains with up to $5$ knots.}
\end{figure}
\end{center}

We simulated chains of size $N$ ranging from $19$ segments to $349$ segments. The results for the number of knots $k$ in the chains is shown in Figure~\ref{fig:knotprob350}. We show the
probability $P(k,N)$ for the occurrence of $k$ knots in a chain of length $N$. The longer the chain is the
more rapidly decreases the probability for having no knot in the chain. This is consistent with the prove
that the probability for having at least one knot in the chain is one if the chain is long enough~\cite{Summers}.
The probability for having exactly one knot in the chain goes through a maximum as do all other probabilities
for fixed $k$. Thus the longer the chain the more likely it is to have several knots in it.

\subsection{Number of the monomers in a knot}

Before we start to investigate the number of monomers in a knot, we have to point out the problems that arise from our knot definition. We
consider a blob of monomers that cannot be completely stretched as a knot. In this definition we assume that the polymer can be completely
stretched! How large is the force which stretches the polymer? We used a force which is high enough to stretch the polymer such that the knot(s)
become tight enough to count only the monomers that really belong to a knot.

In what follows we compare our results to those in the work of Farago, Kantor and Kardar~\cite{bib1}, where knotted polymers were pulled between
two parallel walls. This is a difference to our geometry, but when the polymer is stretched, the interactions between the walls and the polymers
are negligible. Hence the results of\cite{bib1} and those of our investigations are comparable.

Let $n(N)$ be the number of simulated polymers of length $N$, $k$ the number of knots, and $n_k(N)$ the number of simulated chains $N$ with
exactly $k$ knots. $P(k,N)$ is the relative frequency to find $k$ knots in a chain of length $N$. Further let $m_i(N)$ be the number of monomers
participating in the $i$-th simulated polymer of length $N$. Then the number of knot-monomers at fixed length $N$ is

\begin{equation}
M(N)=\sum_{i=1}^{n(N)}m_i(N).
\end{equation}

In\cite{bib1} a power law for the number of knot-monomers is predicted

\begin{equation}
M(N)^{k=1}\cong N^{0.4\pm0.1} \quad .
\end{equation}

For this result only chains with one knot were used. Fitting our data to a power law we obtain an exponent $0.39 \pm0.13$

\begin{equation}
N_m^{k=1}=1.25\cdot N^{0.39\pm0.13} \quad
\end{equation}

\begin{center}
\begin{figure}
\includegraphics[width=\textwidth]{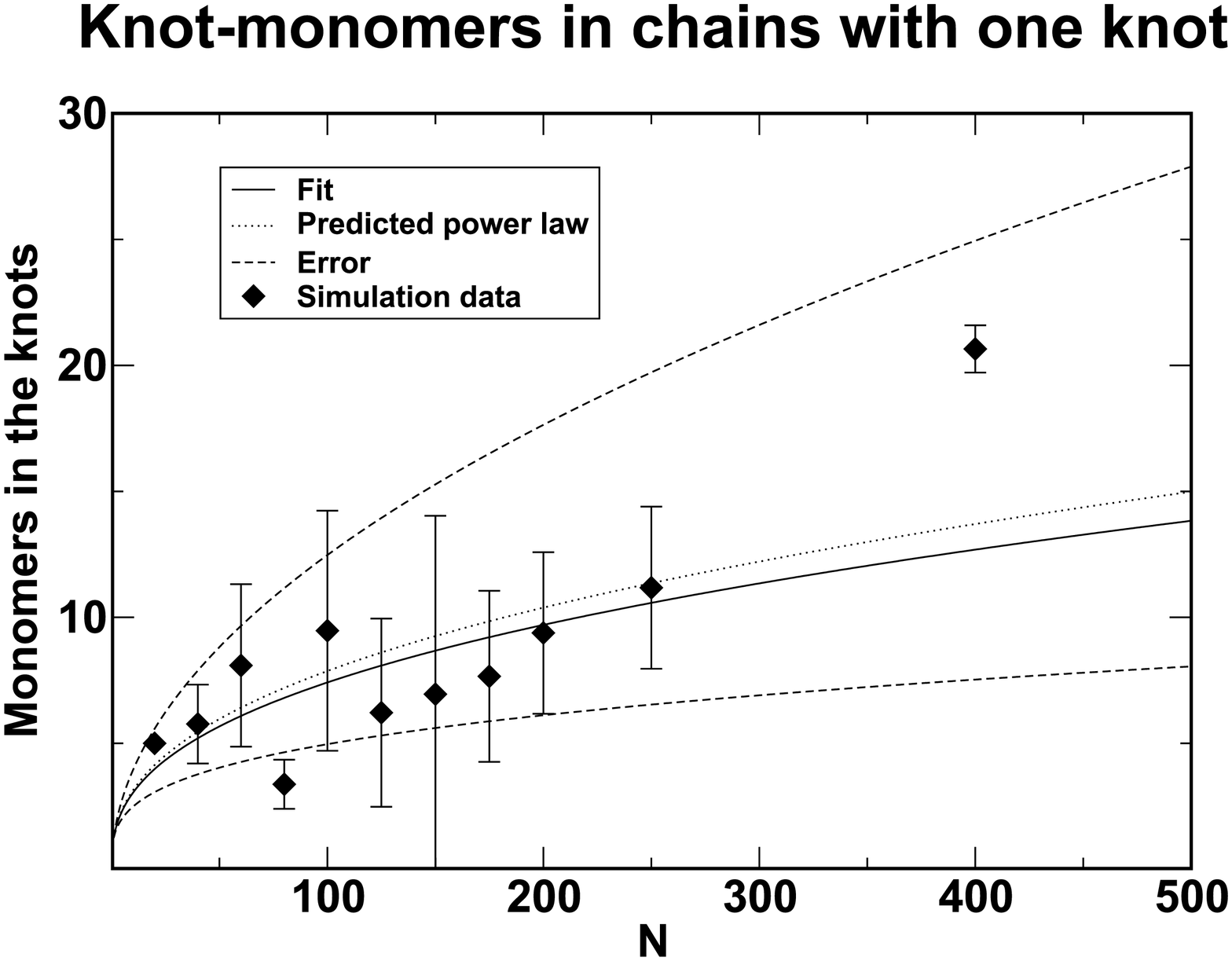}
\caption{\label{fig:knot1mon} Number of monomers in the knots in chains with only one knot. The data as well as a calculations by Farago et. al.
suggest a power law behaviour with an exponent $0.39$.}
\end{figure}
\end{center}

In Figure~\ref{fig:knot1mon} the data is shown. The solid line is the fit to the data and the dotted lines are the power laws with the errors
predicted by\cite{bib1}. The scatter is still very large and especially the last data point seems to be an outlier.

\begin{center}
\begin{figure}[ht]
\includegraphics[width=\textwidth]{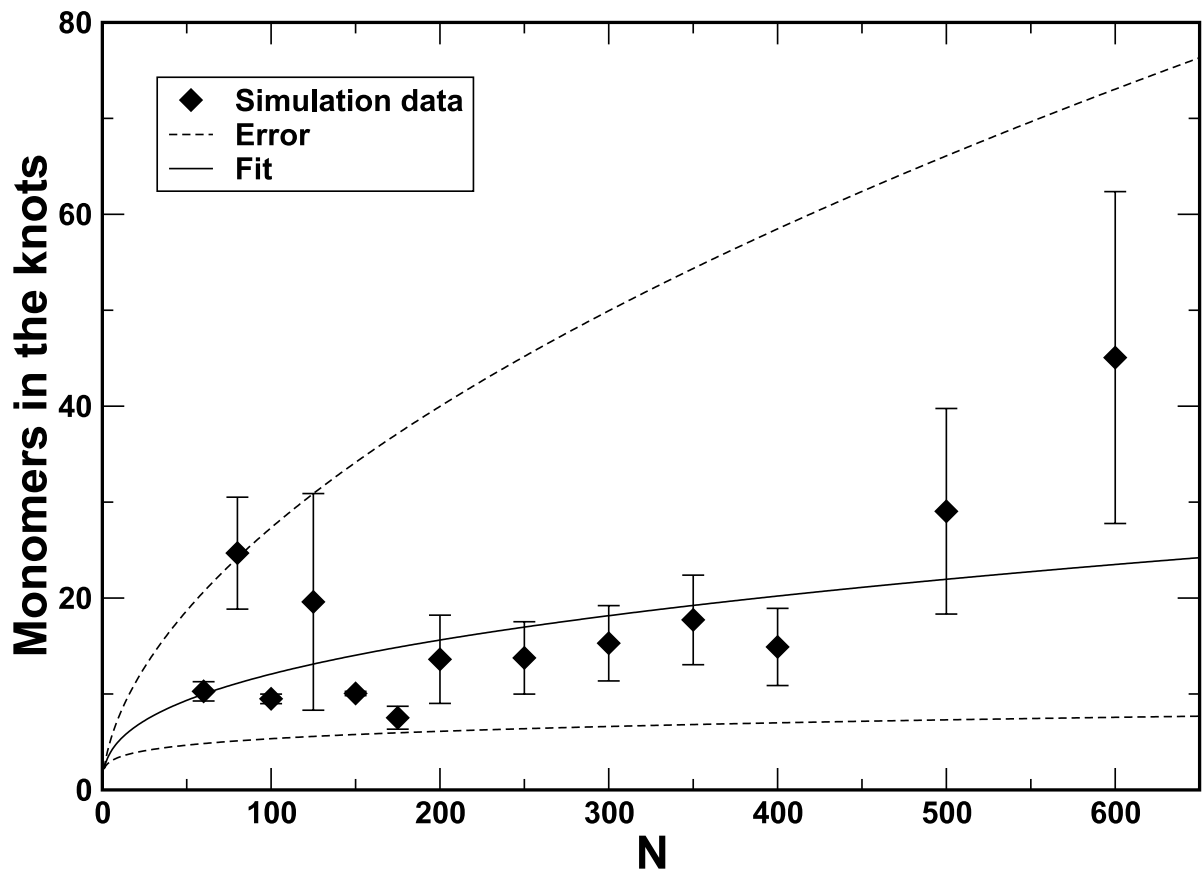}
\caption{ \label{fig:knot2mon}  Number of monomers in the knots in chains with two knots. Again we find a power law behaviour with an exponent }
\end{figure}
\end{center}

If we look at the chains with a number of knots greater than one, we can also fit the data, but only in the two-knot-case reasonable results can
be found. Because of the low number of the highly knotted chains the statistis is not good enough for chains with three or more knots. In Figure~\ref{fig:knot2mon} the results for the two-knot-chains are shown and compared to the $N^{0.4}$ law from\cite{bib1}. The fitted power law we
obtain in this case is

\begin{equation}
N_m^{k=2}=2.18\cdot N^{0.37\pm0.18}.
\end{equation}

This result is still close to the $N^{0.4}$ power law for the one-knot chains.

Averaging over all chains we obtain a stronger dependence on the chain length than in the one or two-knot case. A power law fit yields

\begin{equation}
N_m=0.0058\cdot N^{1.42\pm0.05}.
\end{equation}

\begin{center}
\begin{figure}[ht]
\includegraphics[width=\textwidth]{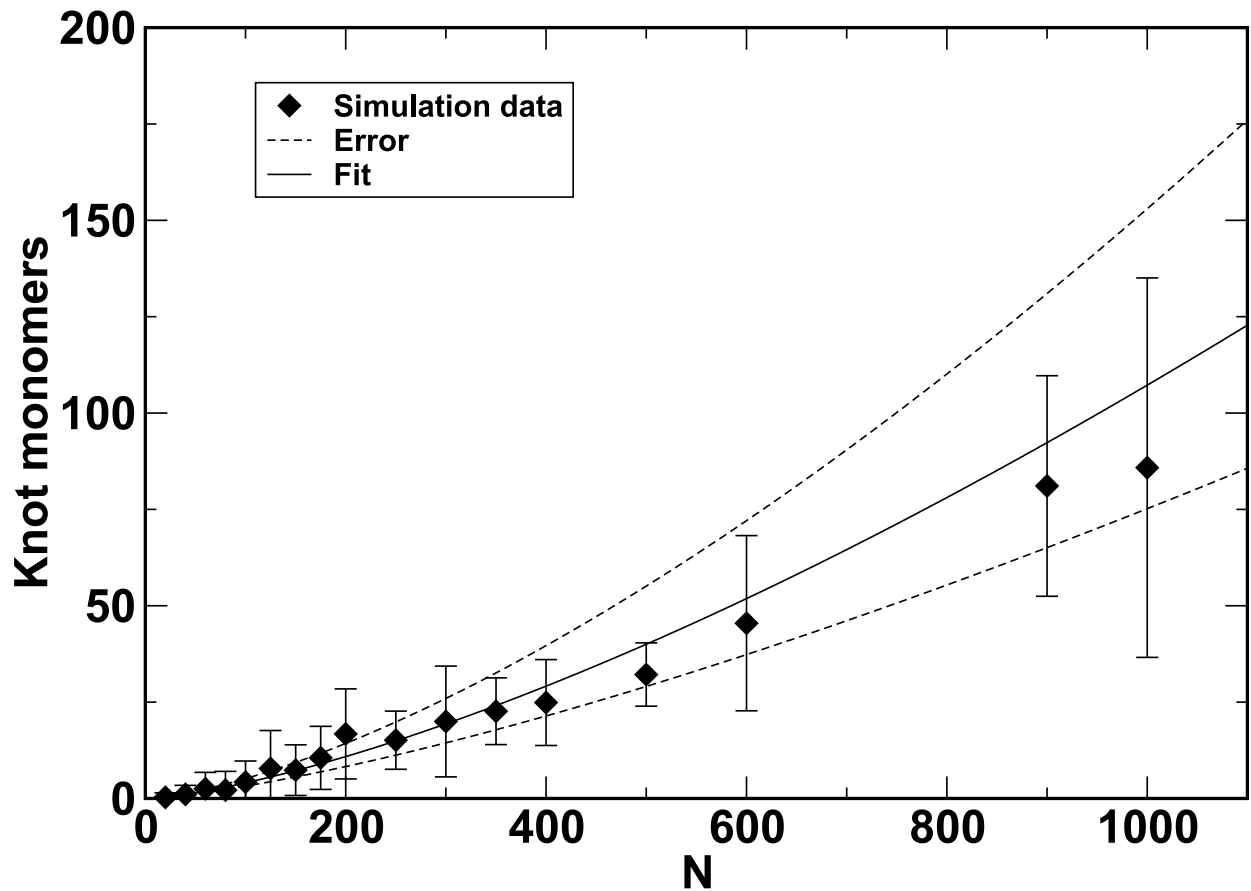}
\caption{\label{fig:knotmon} Number of monomers in the knots. The data suggest a power law behaviour with an exponent $1.42$.}
\end{figure}
\end{center}

This result is shown in Figure~\ref{fig:knotmon} together with the simulation data.

\begin{center}
\begin{figure}[ht]
\includegraphics[width=\textwidth]{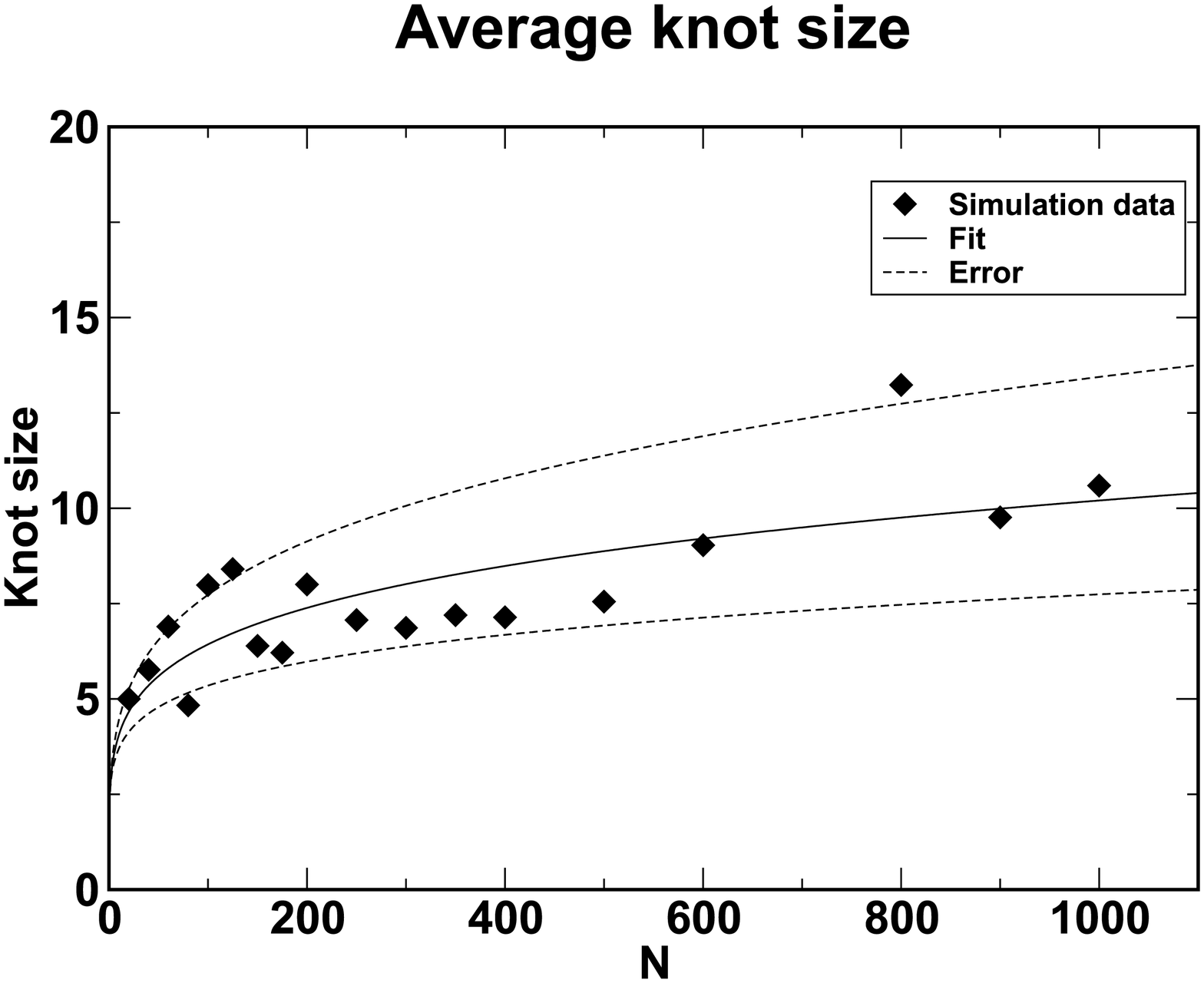}
\caption{\label{fig:monperknot} Average number of monomers in the knots. The plot show a power law fit with an exponent $0.20$.}
\end{figure}
\end{center}

How large are the knots? We averaged over all chains and over all knots (see Figure~\ref{fig:monperknot}). The average value of the knot size
increases rapidly for short chains and saturates. Assuming again a power law behaviour we obtain
\begin{equation}
N_m=2.56\cdot N^{0.20\pm0.04}.
\end{equation}

\section{Discussion}

The average crossing number is one way to characterize knots. Our results show that the topological invariant, which is the basis for the prove
by Diao and coworker is not influenced by excluded volume interactions. Hence it is still unclear, especially in the light of the inconclusive
simulation data, whether the proven law $N\log N + cN$ for the non-excluded volume case changes to a power law, as suggested by
Buck~\cite{Buck}. Much larger chains are needed to clearly discern between the possibilities. A rough estimate shows that we need a decade
longer chains. Here the problem is that despite the well developed Pivot algorithm the computation of the excluded volume interaction is so
time-consuming that for now it seems to be possible to do such a calculation only expending a truly fair amount of computing resources.

The statistics of knots show a power law behaviour for all of the quantities investigated in this paper. Here once again the gathering of
sufficient statistics for the longer chains is difficult to improve. Desirable would be to increase the chain length by a factor of ten to give
a good estimate on the corresponding exponents. What is also lacking is a good derivation of the power laws in one framework.

\begin{acknowledgments}

We are very grateful to J. Odenheimer and K. Binder for discussions.
\end{acknowledgments}


\newpage

\newpage

\newpage

\newpage

\newpage

\newpage

\newpage

\newpage

\newpage

\newpage


\end{document}